              [1999/12/01 v1.4c Il Nuovo Cimento]
\documentclass{cimento}

\def\bb{\bibitem}

\def\rp#1#2{{#1\over#2}}
\def\bar{\begin{eqnarray}}
\def\ear{\end{eqnarray}}
\def\eqi{\begin{equation}}
\def\eqf{\end{equation}}
\def\derp#1#2{\rp{\partial{#1}}{\partial{#2}}}
\def\gi#1#2{g_{{#1}{#2}}}

\def\dert#1#2{\frac{d#1}{d#2}}

\def\rfr#1{eq.(\ref{#1})}
\def\rfrs#1#2{eqs.(\ref{#1})-(\ref{#2})}
\def\Rfr#1{Eq.(\ref{#1})}

\def\ct#1{\cite{#1}}
\def\lb#1{\label{#1}}

\title{Light deflection and time delay in the gravitational field
of a spinning body }
\author{Lorenzo~Iorio\from{ins:x}}
\instlist{\inst{ins:x} Dipartimento Interateneo di Fisica dell'
Universit{\`{a}} di Bari, Via Amendola 173, 70126, Bari}
\PACSes{\PACSit{95.30}{Relativity and gravitation}
        \PACSit{04.20}{Fundamental problems and general formalism}}
\begin{document}

\maketitle

\begin{abstract}
In this paper we consider the possibility of measuring the
corrections induced by the square of the parameter $a_g$ of the
Kerr metric to the general relativistic deflection of
electromagnetic waves and time delay in an Earth based experiment.
It turns out that at laboratory scale the rotational effects
exceed definitely the gravitoelectric ones which are totally
negligible. By using a small rapidly rotating sphere as
gravitating source on the Earth the deflection of a grazing light
ray amounts to $10^{-13}$ rad and the time delay is proportional
to $10^{-23}$ s. These figures are determined by the upper limit
in the attainable values of $a_g$ due to the need of preventing
the body from exploding under the action of the centrifugal
forces. Possible criticisms to the use of the Kerr metric at
$a_g^2$ level are discussed.
\end{abstract}
%--------------------------------------------------------------
\section{Introduction}
The effect of the proper angular momentum $J$ of a body of mass
$M$, assumed as source of the gravitational field, is accounted
for, in General Relativity, by the parameter $a_g=\rp{J}{Mc}$,
where $c$ is the speed of light in vacuum, entering, e.g., the
Kerr metric. It has the dimensions of a length and is of order
$\mathcal{O}(c^{-1})$. The other characteristic length is the
Schwarzschild radius $R_s=\rp{2GM}{c^2}$, where $G$ is the
Newtonian gravitational constant, which enters both Schwarzschild
and Kerr metrics.

Traditionally, in almost all the applications of the Kerr metric
to the motion of test particles or electromagnetic waves in the
space--time of a central spinning mass the square of $a_g$, which
is of order $\mathcal{O}(c^{-2})$ and enters the diagonal
components of the metric tensor, is neglected. However, in a
recent stimulating letter in ref. \ct{ref:tart02} it is suggested
that, instead, it would be better to account for it in view of
possible experimental setups on the Earth. Indeed, at laboratory
scale, while the effects due to $R_s$, of order
$\mathcal{O}(c^{-2})$, are certainly negligible because of the
smallness of the product $GM$, those due to $a_g^2$, of order
$\mathcal{O}(c^{-2})$ as well, could become interesting because,
for certain symmetric geometries of the source body, it depends
only on the fourth power of its radius $R$ and the square of its
angular velocity $\Omega$.

In this paper we want to explore this scenario by working out the
contributions of $a_g^2$ to two classical tests of General
Relativity: the deflection of the electromagnetic waves and the
time delay \ct{ref:ciuwhe95}. We will consider the motion of
photons in the equatorial plane of a central rotating body at
rest.
%in the context of the Post--Newtonian
%approach.
In a number of papers these general relativistic effects have been
worked out for various systems of gravitating bodies endowed with
mass--monopole, spin--dipole and time--dependent mass and spin
multipoles for different states of motions both of the source of
electromagnetic waves and the observer \ct{ref:kop97,
ref:blanetal01, ref:kopmash, ref:ser02, ref:sercard02,
ref:ciufricc02} by means of a variety of general mathematical
approaches, but the influence of $a_g^2$ has never been
considered.

A possible objection to the present calculations could be that
they are based on the use of the Kerr metric in representing the
gravitational field of an extended object at the $a_g^2$ order.
Indeed, it should be reminded that, up to now, nobody has been
successful in extending the Kerr metric from the empty space to
the interior of a body. In fact, the field of a real rotating body
would be endowed of various multipolar mass moments, etc. and if
we do not know what is the matter source, as in the case of the
Kerr metric, we do not know how to cope with them\footnote{
However, in the present case we could argue that, whatever the
description of the non--sphericity of the central mass could be,
it should not affect in any relevant way the obtained results, as
confirmed a posteriori by the presented calculations.}. Then, one
could wonder if, at the level of approximation considered, the
Kerr metric and the derived consequences have physical meaning.
However, notice that in ref. \ct{ref:tart98} the author use the
full Kerr metric in deriving the general relativistic corrections
of the Sagnac effect. In ref. \ct{ref:tartrug02a} the Kerr metric
is used in order to sketch a possible Earth based laboratory
experiment aimed to testing gravitomagnetism at order $a_g$.
Moreover, recently some efforts are or will be directed (M.
McCallum, private communication, 2002) towards possible extensions
of the Kerr metric to various interior matter sources. Many
efforts have been dedicated to the construction of sources which
could represent some plausible models of, e.g., stars. In ref.
\ct{ref:harttorn68} the authors have tried to connect the external
Kerr metric to the multipolar structure of various types of stars
assumed to be rigidly and slowly rotating. Terms of greater than
the second order in the angular velocity were neglected. In ref.
\ct{ref:ali01} a physically reasonable fluid source for the Kerr
metric has been obtained. For previous attempts to construct
sources for the Kerr metric see refs.
\ct{ref:haggmar81,ref:hagg90}. In ref.\ct{ref:wilt03} a general
class of solutions of Einstein's equations for a slowly rotating
fluid source, with supporting internal pressure, is matched to the
Kerr metric up to and including first order terms in angular speed
parameter. So, in this context, our calculations could be useful
in order to give an idea of what could happen in this case and,
more generally, if experiments testing different scenarios at
order $\mathcal{O}(J^2)$ are realistically conceivable.

A possible subject for further analysis could be an investigation
of the correspondence with the Post--Newtonian expansion at the
$J^2$ level. At the first order in $J$ the Post--Newtonian
expansion accounts for the angular momentum of the source in the
off--diagonal components $g_{0i}\ i=1,2,3$ via a $c^{-3}$
term\footnote{This may explain the fact that Kerr metric is
commonly accepted in describing the Lense--Thirring effect and the
gravitomagnetic clock effects which are linear in $J$. }. So, at a
first glance, it could be guessed that the Post--Newtonian
expansion should account for $J^2$ with a $c^{-6}$ term which,
instead, is absent in the Kerr metric. Indeed, as we will see, our
effects depends only on $a_g^2$.

Recently, in ref. \ct{ref:tart03} an approximated solution of the
Einstein field equations for a rotating, weakly gravitating body
has been found. It seems promising because both external and
internal metric tensors have been consistently found, together an
appropriate source tensor. Moreover, the mass and the angular
momentum per unit mass are assumed to be such that the mass
effects are negligible with respect to the rotation effects. As a
consequence, only quadratic terms in the body's angular velocity
has been retained. One main concern is that it is not clear if
there is a real mass distribution able to generate the found
source tensor.

As can be seen, the subject seems to be rather open and, in the
author's opinion, worth of investigation, at least in order to get
some estimates of the orders of magnitude involved.

 The paper is organized as follows: in section 2 we
will derive the geodesic motion of a test particle of mass $m$ in
the equatorial plane of a rotating body. In section 3 we will
specialize it to the photons and will work out the deflection of
electromagnetic waves due to $a_g^2$. In section 4 its effect on
the time delay is examined. Section 5 is devoted to the discussion
of the obtained results and to possible applications to Earth
laboratory experimental scenarios.
%---------------------------------------------------------------
\section{Plane geodesics in the Kerr metric}
Let us consider,
for the sake of concreteness, a spherically symmetric rigid body
of mass $M$, radius $R$ and proper angular momentum $J$ directed
along the $z$ axis of an asymptotically inertial frame
$K\{x,y,z\}$ whose origin is located at the center of mass of the
body. By adopting, as usual, the coordinates \bar
x^{0} & = & ct,\\
x^{1} & = & r,\\
x^{2} & = & \theta,\\
x^{3} & = & \phi,\ \ear the components of the Kerr metric tensor
are \ct{ref:boli67}\bar
\gi00 & = & 1-\frac{R_s r}{\varrho^2}, \\
\gi11 & = & -\frac{\varrho^2}{\Delta},\\
\gi22 & = & -\varrho^2,\\
\gi33 & = & -\sin^2\theta\left[r^2+a_g^2+\frac{R_s r}{\varrho^2} a_g^2\sin^2\theta\right],\\
\gi03 & = & \frac{R_s r}{\varrho^2} a_g\sin^2\theta, \ear
with \bar\varrho^{2} & = & r^{2}+a_g^{2}\cos^{2}\theta,\\
\Delta & = & r^{2}-R_s r+a_g^{2}. \ear For a spherical body
spinning at angular velocity $\Omega$ and with moment of inertia
$I=\rp{2}{5}MR^2$ the characteristic length $a_g$ becomes \eqi
a_g=\rp{I\Omega}{Mc}=\rp{2}{5}\rp{R^2\Omega}{c}.\eqf

It is important to note that neither the Newtonian gravitational
constant $G$ nor the mass $M$ are present in $a_g$ which is
determined only by the geometrical and kinematical properties of
the body. This is a feature which will turn out to be very
relevant in proposing experiments at laboratory scale.

By putting \bar\varepsilon &=&\frac{R_s}{r}\\
\alpha &=& \frac{a_g}{r},\ear the line element for a test particle
of mass $m$ moving in the space--time of the central body in its
equatorial plane, i.e. at fixed $\theta=\rp{\pi}{2}$, is given by
\ct{ref:chan83} \eqi c^2=(1-\varepsilon)(\dot x^0)^2-\frac{(\dot r
)^2}{1-\varepsilon+\alpha^2}-r^2(1+\alpha^2+\varepsilon\alpha^2)(\dot\phi)^2+2\varepsilon\alpha
r\dot x^0\dot \phi.\lb{ds}\eqf In \rfr{ds} the overdot denotes the
derivative with respect to proper time $\tau$.
%In view of an
%application to possible experimental scenarios at laboratory
%scale, it is very instructive to
%note that \bar\alpha^2&\sim&\rp{1}{c^2}\\
%\varepsilon&\sim&\rp{GM}{c^2}\\
%\varepsilon^2&\sim&\rp{G^2M^2}{c^4}\\
%\varepsilon\alpha&\sim&\rp{GM}{c^3}\\
%\varepsilon\alpha^2&\sim&\rp{GM}{c^4}.\ear This implies that, in
%general, for the symmetry of the considered problem, the terms
%with the square of the characteristic length $a_g$ may give not
%negligible contributions.

From the Lagrangian \eqi\mathcal{L}=\rp{m}{2} g_{\mu\nu}\dot
x^{\mu}\dot x^{\nu}=\rp{m}{2}\left[(1-\varepsilon)(\dot
x^0)^2-\frac{(\dot r
)^2}{1-\varepsilon+\alpha^2}-r^2(1+\alpha^2+\varepsilon\alpha^2)(\dot\phi)^2+2\varepsilon\alpha
r\dot x^0\dot \phi\right],\eqf since the field is stationary and
axially symmetric, it is possible to obtain the following
constants of motion \bar
\derp{\mathcal{L}}{\dot x^0} &=& m(1-\varepsilon)\dot x^0+m\varepsilon\alpha r\dot\phi=K\lb{kappa}\\
\derp{\mathcal{L}}{\dot \phi}
&=&-mr^2(1+\alpha^2+\varepsilon\alpha^2)\dot\phi+m\varepsilon\alpha
r\dot x^0=H.\lb{acca} \ear Let us write \rfr{ds} as
\eqi\left(\rp{c}{\dot\phi}\right)^2=(1-\varepsilon)\left(\rp{\dot
x^0
}{\dot\phi}\right)^2-\frac{1}{1-\varepsilon+\alpha^2}\left(\rp{\dot
r
}{\dot\phi}\right)^2-r^2(1+\alpha^2+\varepsilon\alpha^2)+2\varepsilon\alpha
r\left(\rp{\dot x^0}{\dot\phi}\right). \lb{dds}\eqf

In view of an application to possible experimental scenarios at
laboratory scale, it is very instructive to
note that \bar\alpha^2&\sim&\rp{1}{c^2}\\
\varepsilon&\sim&\rp{GM}{c^2}\\
\varepsilon^2&\sim&\rp{G^2M^2}{c^4}\\
\varepsilon\alpha&\sim&\rp{GM}{c^3}\\
\varepsilon\alpha^2&\sim&\rp{GM}{c^4}.\ear This implies that, at
laboratory scale, the terms with the square of the characteristic
length $a_g$ may give not negligible contributions with respect to
the other terms which are quite negligible because of the presence
of $GM$ and/or because they are of order $\mathcal{O}(c^{-n})$,
$n\geq 3$.

From \rfrs{kappa}{acca}, by keeping, in a first step, only terms
of order $\mathcal{O}(c^{-2})$, it is possible to obtain
\bar\left(\rp{\dot x^0 }{\dot\phi}\right) & = &
\rp{K}{H}\rp{r^2(\alpha^2+1)}{(\varepsilon-1)}\lb{td}\\
\rp{1}{\dot\phi} &=& \rp{m}{H}\rp{r^2(1+\alpha^2-\varepsilon
)}{(\varepsilon-1)}.\lb{fid}\ear With \bar u &\equiv&\rp{1}{r}\\
\left(\rp{\dot r}{\dot
\phi}\right)&=&\dert{r}{\phi}=-\rp{1}{u^2}\dert{u}{\phi}\ear and
by inserting \rfrs{td}{fid} in \rfr{dds} one obtains, to order
$\mathcal{O}(c^{-2})$
\eqi\left(\dert{u}{\phi}\right)^2=\left(\frac{K^2-m^2c^2}{H^2}\right)-u^2+\rp{m^2
c^2}{H^2}R_s u-3\rp{m^2 c^2}{H^2}a_g^2
u^2+3\left(\rp{K}{H}\right)^2 a_g^2 u^2 +R_s u^3 -2a_g^2
u^4.\lb{dudf1}\eqf By taking the derivative of \rfr{dudf1} it can
be obtained \eqi\rp{d^2
u}{d\phi^2}=-u+\rp{m^2c^2R_s}{2H^2}-3\rp{m^2c^2}{H^2}a_g^2 u+
3\left(\rp{K}{H}\right) a_g^2 u+\rp{3}{2}R_s u^2-4a_g^2
u^3.\lb{dudf2}\eqf Note that in \rfr{dudf2} the first three terms
of the right hand side are of order $\mathcal{O}(c^0)$ and the
following three terms are of order $\mathcal{O}(c^{-2})$.
%--------------------------------------------------------------------------------
\section{The deflection of light}
In order to cope with the case of photons having zero rest mass,
let us pose $m=0$ in \rfrs{dudf1}{dudf2} so to obtain \bar
\left(\dert{u}{\phi}\right)^2 & = &
\left(\frac{K}{H}\right)^2-u^2+3\left(\rp{K}{H}\right)^2 a_g^2 u^2
+R_s u^3
-2a_g^2 u^4,\lb{ddudf1}\\
\rp{d^2 u}{d\phi^2} & = & -u + 3\left(\rp{K}{H}\right)^2 a_g^2
u+\rp{3}{2}R_s u^2-4a_g^2 u^3. \lb{ddudf2}\ear

In the following, without loss of generality, we will assume to
count the angle $\phi$ from the point of closest approach so that
$u(\phi=0)=u_{max}=\rp{1}{r_{min}}\equiv\rp{1}{b}$.
%where $b$
%defines the impact parameter.
%-----------------------------------------------------------
\subsection{The effect of ${a}_{\bf g}^{\bf 2}$}
%------------------------------------------------------------
In order to calculate the effect of $a_g^2$, which is of order
$\mathcal{O}(c^{-2})$, let us drop the terms in $R_s$  in
\rfrs{ddudf1}{ddudf2}: they become\footnote{This approximation is
fully satisfied for laboratory scale bodies, as it will become
clear later.} \bar \left(\dert{u}{\phi}\right)^2 & = &
\left(\frac{K}{H}\right)^2-u^2+3\left(\rp{K}{H}\right)^2 a_g^2 u^2
-2a_g^2 u^4,\lb{ddudff1}\\
\rp{d^2 u}{d\phi^2} & = & -u + 3\left(\rp{K}{H}\right)^2 a_g^2
u-4a_g^2 u^3. \lb{ddudff2}\ear By evaluating \rfr{ddudff1} for
$\phi=0$ it is possible to obtain $K/H$. Indeed, at the closest
approach we have $u=u_{max}$ and $\dert{u}{\phi}=0$, so that, from
\rfr{ddudff1} at order $\mathcal{O}(c^0)$
\eqi\left(\rp{K}{H}\right)^2\sim u_{max}^2.\eqf Then,
\rfr{ddudff2} becomes \eqi \rp{d^2 u}{d\phi^2} + (1 - 3u_{max}^2
a_g^2) u +4a_g^2 u^3=0. \lb{dddudf2}\eqf In order to solve this
equation, it is useful to pose \bar
\omega_0^2 &\equiv& (1 - 3u_{max}^2 a_g^2),\\
\gamma &\equiv& \rp{4a_g^2}{(1 - 3u_{max}^2 a_g^2)}\sim
4a_g^2.\ear This allows to write \rfr{dddudf2} as \eqi \rp{d^2
u}{d\phi^2} + \omega^2_0 u +\gamma\omega_0^2 u^3=0.
\lb{anarmo}\eqf The solution of an equation of the form of
\rfr{anarmo} is given by \ct{ref:migu83}\eqi
u=u_{max}\left(1-\gamma
\rp{u_{max}^2}{32}\right)\cos\omega\phi+\rp{\gamma
u_{max}^3}{32}\cos3\omega\phi,\lb{soluz}\eqf with
\eqi\omega^2=\rp{\omega^2_0}{1-\rp{3}{4}\gamma
u_{max}^2}.\lb{omega}\eqf At order $\mathcal{O}(c^{-2})$ it reads
\eqi\omega^2\sim 1;\eqf then \rfr{soluz} becomes \eqi
u=u_{max}\left(1-\gamma
\rp{u_{max}^2}{32}\right)\cos\phi+\rp{\gamma
u_{max}^3}{32}\cos3\phi.\lb{soluzi}\eqf From \rfr{soluzi} it can
be easily obtained the deflection angle. Indeed, let us assume
that the photon flies away at infinity, i.e. $u\rightarrow 0$, not
for $\overline{\phi}=\rp{\pi}{2}$, as it should be the case for
the straight line propagation in flat Minkowskian space--time, but
for ${\overline{\phi}}^{'} =\rp{\pi}{2}+\delta_{a_g^2}$. So,
\rfr{soluzi} yields \eqi \delta_{a_g^2}=-\rp{\gamma
u_{max}^2}{32}\left(1+\rp{\gamma u_{max}^2
}{32}\right)\sim-\rp{\gamma
u_{max}^2}{32}\sim-\rp{a_g^2}{8b^2}=-\rp{J^2}{8 M^2 c^2
b^2}=-\rp{1}{50}\rp{R^4\Omega^2}{c^2 b^2}.\lb{angolo}\eqf
%---------------------------------------------------
\section{The time delay}
The coordinate time interval between the emission and the
reception of an electromagnetic signal at two different points $A$
and $B$ (one-way travel) can be written as \ct{ref:lanlif79} \eqi
d t=\rp{\sqrt{(\gi0i \gi0j-\gi{i}j\gi00)dx^i
dx^j}}{cg_{00}}.\lb{dt}\eqf In it the Latin indices run from 1 to
3 and the Einstein summation convention is adopted. For the Kerr
metric at $\theta=\rp{\pi}{2}$ \rfr{dt} becomes \eqi d
t=\rp{\sqrt{({\gi03}^2 -\gi33\gi00)d\phi^2-\gi11\gi00
dr^2}}{cg_{00}}\sim\rp{\sqrt{-\gi33 d\phi^2-\gi11
dr^2}}{c\sqrt{g_{00}}}\sim\rp{\sqrt{r^2(1+\alpha^2)
d\phi^2+\rp{dr^2}{1-\varepsilon+\alpha^2}}}{c\sqrt{1-\varepsilon}}.\lb{dtt}\eqf
\Rfr{dtt} has been obtained by neglecting $\gi03$, proportional to
$\varepsilon\alpha$, and $\varepsilon\alpha^2$ in $\gi33$. By
assuming the polar axis as $x$ axis parallel to the (almost)
straight line motion of the photons, we can neglect $r^2d\phi^2$
with respect to $dr^2$ and we can pose $dr\sim dx,\
r\sim\sqrt{x^2+b^2}$. With these approximation and by neglecting,
as usual, the terms proportional to $\varepsilon^2$ and
$\varepsilon\alpha^2$, \rfr{dtt} yields \eqi d
t=\left[1+\rp{R_s}{\sqrt{x^2+b^2}}-\rp{a_g^2}{2(x^2+b^2)}\right]\rp{dx}{c}.\lb{ddtt}\eqf
By integrating \rfr{ddtt} from $x_A$ to $x_B$ it can be obtained
\eqi \Delta t=\Delta t_0+\Delta t_{\rm GE}+\Delta t_{a_g^2}\eqf
with
\bar \Delta t_0 & = & \rp{x_B-x_A}{c},\lb{uno}\\
\Delta t_{\rm GE} & = & \rp{R_s}{c}\ln\left|
\rp{x_B+\sqrt{{x_B}^2+b^2}}
{x_A+\sqrt{{x_A}^2+b^2}}\right|,\lb{due}\\
\Delta t_{a_g^2}& = & -\rp{a_g^2}{2cb}\left[\arctan
\left(\rp{x_B}{b}\right)-\arctan \left(\rp{x_A}{b}\right)
\right]\lb{tre}.\ear \Rfr{uno} is the ordinary time interval in
the flat Minkowski space--time. \Rfr{due} is the well known
gravitoelectric Shapiro time delay \ct{ref:shap64} due to the
spherical mass $M$ supposed non--rotating. \Rfr{tre} represents
the new term due to $a_g^2$ of the Kerr metric.
%---------------------------------------------------
\section{Discussion}
Here we investigate the two relativistic effects worked out and
examine the possibility of measure them in terrestrial
experiments\footnote{It is important to notice that in an
astrometric scenario, e.g., at the limb of the Sun or of a compact
star, the gravitoelectric term of order $\mathcal{O}(c^{-2})$ and
the gravitomagnetic term of order $\mathcal{O}(c^{-3})$ are far
larger than the terms due to $a_g^2$ and the approximations used
until now are no longer valid. See, e.g., the refs.
\ct{ref:ricmat82a, ref:ricmat82b, ref:ricmat83,
ref:bertgiamp92}.}.
\subsection{The deflection of light}
\Rfr{angolo} shows many interesting features
\begin{itemize}
  \item  It is of order $\mathcal{O}(c^{-2})$, as the well known
  gravitoelectric deflection angle \ct{ref:ciuwhe95} \eqi\delta_{\rm GE}=2\rp{R_s}{b}=4\rp{GM}{c^2
  b}\eqf
  \item It is independent of the frequency of the electromagnetic
  radiation, as $\delta_{\rm GE}$
  \item It is opposite in sign with respect to $\delta_{\rm GE}$
  acting as a diverging effect
  \item  It does not contain neither the Newtonian
  gravitational constant $G$ nor the mass $M$ of the central body,
  a feature that may be of great help in an Earth laboratory experiment
  \item It depends on $b^{-2}$, contrary to $\delta_{\rm GE}$
  which, instead, depends on $b^{-1}$.
  \item  It is insensitive to the sense of rotation of the
  central body: it is a pity since such a feature could have been
  useful in order to generate some particular signature.
\end{itemize}

Let us investigate what could happen in a possible Earth
laboratory experiment. E.g., if we assume as central body a small
sphere of 2.5 cm radius, mass of 111 g and spinning at\footnote{
Indeed, according to ref. \ct{ref:tart02b}, the maximum peripheral
linear speed attainable by a spinning body is $v_{max}\sim 10^3$ m
s$^{-1}$. } $4\times 10^4$ rad s$^{-1}$ we would have \bar
R_s^{\rm sphere} &=& 1.647\times 10^{-28}\
\textrm{m},\\
a_g^{\rm sphere} &=& 3.61\times 10^{-8}\ \textrm{m}. \ear For a
grazing light ray, which means that
\eqi\delta_{a_g^2}=-\rp{1}{50}\left(\rp{\Omega
R}{c}\right)^2=-\rp{1}{50}\left(\rp{v_{\rm per}}{c}\right)^2,\eqf
the deflection would amount to \bar \delta_{\rm GE}^{\rm sphere} &
= &1.31\times 10^{-26}\ \textrm{rad}=2.717\times 10^{-21}\
\textrm{asec},\\
\delta_{a_g^2}^{\rm sphere} & = & -2.606\times 10^{-13}\
\textrm{rad}=-5.37\times 10^{-2} \ \mu\textrm{asec}.\ear In this
case the deflection due to $a_g^2$ is 13 orders of magnitude
larger than the gravitoelectric deflection.
%---------------------------------------------------
\subsection{The time delay}
\Rfr{tre} exhibits the following characteristics
\begin{itemize}
  \item  It is of order $\mathcal{O}(c^{-3})$, as $\Delta t_{\rm GE}$
  \item It is independent of the frequency of the electromagnetic
  radiation, as $\Delta t_{\rm GE}$
  \item  It does not contain neither the Newtonian
  gravitational constant $G$ nor the mass $M$ of the central body,
  contrary to $\Delta t_{\rm GE}$
  \item Its amplitude depends on $b^{-1}$, contrary to $\Delta t_{\rm GE}$ which
  depends only on the characteristic length $R_s$.
  \item  It is insensitive to the sense of rotation of the
  central body
\end{itemize}

At laboratory scale, for the same sphere as before, we have \bar
\Delta t ^{\rm
sphere}_{\rm GE} & \propto & 5.49\times 10^{-37}\ {\rm s },\\
\lb{tdgel} \Delta t ^{\rm sphere}_{a_g^2} & \propto & -8.7\times
10^{-23}\ {\rm s }.\lb{dtgml}\ear Also for the time delay the
correction due to $a_g^2$ is 14 orders of magnitude larger than
the gravitoelectric one. However, the detection of a time
difference of the order of $10^{-23}$ s is presently out of
discussion: suffice it to say that $10^{-23}$ s is the typical
lifetime of a strongly decaying particle (e.g. $\rho$ or $N^*$).
%----------------------------------------
\subsection{Conclusion}
In this paper we have calculated the effect of the square of the
parameter $a_g$ entering the Kerr metric on the general
relativistic deflection of electromagnetic waves and the time
delay in the gravitational field of a central rotating source at
rest.

Then, we have analysed those two effects in a possible
experimental scenario in an Earth based laboratory with a small
spinning sphere.  Due to their extreme smallness, we have
neglected the terms involving $GM$. The maximum attainable linear
velocity $v_{\rm max}\sim 10^3 $ m s$^{-1}$, due to the effects of
the centrifugal forces, puts severe upper constraints to the
magnitude of $a_g$ in a laboratory experiment. The derived effects
are to small to be detected.

The calculations performed here loose their validity in an
astrophysical context because, in this case, neglecting the other
terms proportional to $GM$ with respect to $a_g^2$ is no more
allowed.
\section*{Acknowledgements}
I wish to thank S.M. Kopeikin, A. Tartaglia, M.L. Ruggiero and M.
MacCallum for useful remarks and insightful discussions. I am
grateful also to B. Mashhoon for the references kindly given me.
Thanks also to L. Guerriero for his supporting me in Bari.
%-----------------------------------------

\end{document}